\documentclass[preprint,showpacs,preprintnumbers,amsmath,amssymb]{revtex4}

\usepackage{graphicx}
\usepackage{dcolumn}
\usepackage{bm}

\begin{document}


\title{Finite width induced modification to the electromagnetic form factors of spin-1 particles.}

\author{D. Garc\'{\i}a Gudi\~no}
\affiliation{Instituto de F\'{\i}sica,  Universidad Nacional Aut\'onoma de M\'exico, AP 20-364,  M\'exico D.F. 01000, M\'exico}%
\author{G. Toledo S\' anchez}%
\affiliation{Instituto de F\'{\i}sica,  Universidad Nacional Aut\'onoma de M\'exico, AP 20-364,  M\'exico D.F. 01000, M\'exico}
\affiliation{Department of Physics, Florida State University, Tallahasse, FL 32306}


\date{\today}

\begin{abstract}
The inclusion of the unstable features of a spin-1 particle,  without breaking the electromagnetic gauge invariance, can be properly accomplished by including higher order contributions as done in the so-called fermion loop scheme (for the $W$ gauge boson), and the boson loop scheme (for vector mesons).  This induces a non trivial modification to the electromagnetic vertex of the particle, which must be considered in addition to any other contribution computed as stable particles.
 Considering the modified electromagnetic vertex, we obtain general expressions for the corresponding corrections to the multipoles as a function of the mass of the particles in the loop. For the $W$ gauge boson no substantial deviations from the stable case is observed. For the $\rho$ and $K^*$ mesons the mass of the particles in the loop makes a significant effect, and can be comparable with corrections of different nature .
 \end{abstract}


\maketitle

\section{Introduction}
The electromagnetic properties of spin-1 particles ($V$) can help us to understand the symmetry structure and interactions of the fundamental particles. For example, considering $V$ as a stable state, the electromagnetic properties of the $W$ gauge boson  are predicted by the symmetry structure of the standard model, while for the $\rho$ meson several predictions exist, based on effective models of the strong interaction binding the quarks \cite{hawes, cardarelli,melo,ho,maris}, and from lattice-QCD \cite{hed,lee}. 
However, they are not stable states.  Therefore, in order to draw definite conclusions, a complete study of the {\em additional} effects due to their instability is mandatory.

The proper theoretical description of such states requires to incorporate their unstable features (parameterized by their finite decay width, $\Gamma$) in an electromagnetic gauge invariant way. To do so, several schemes have been developed, we can mention, for example, the so called fermion loop scheme \cite{baur95,argyres95,beuthe} and the boson loop scheme \cite{bls} (suitable for the $W$ and $\rho$  bosons respectively), which consider that $\Gamma$ is naturally included in the calculations by taking into account the absortive contributions in the electromagnetic vertex and in the propagator. 
Under these schemes, the electromagnetic vertex $VV\gamma$ is modified respect to the tree level form in a non-trivial way. This implies that the electromagnetic structure itself suffers modifications.

 In the present work we analyze the most general results for the electromagnetic vertices obtained in the loop schemes, which include the mass of the particles in the loops, and extract the expressions for the modified form factors. In order to exhibit the size of the corrections due to the unstable nature of the particles, we compare our results for the magnetic dipole moment (MDM) and electric quadrupole moments of the $W$, $\rho $ and $K^*$ mesons with others computed in the literature for contributions from different nature. The gauge invariance requirement, along with the modification in the propagator inherent to the schemes, allows to identify the proper complex renormalization factor of the vector field, which keep the electric charge free of radiative correction contributions.
 
\section{Finite width effects}
The schemes developed in \cite{baur95,argyres95,beuthe,bls} for the introduction of the finite width effects, while keeping electromagnetic gauge invariance, are based in two main observations:  In quantum field theory the width is naturally included in the imaginary part of the self-energy of the particles and, the Ward identity is respected
at all orders in perturbation theory. These facts are exploited in those schemes by including the resummation of the fermion/boson loops in the propagator and the corrections in the electromagnetic vertices. Then, the imaginary part of the fermion/boson loops introduces the tree level width in the gauge boson propagator and, the gauge invariance is not violated since the fermion/boson loops obey the Ward identity order by order.

At tree level, the propagator for a vector boson of mass $M_V$ can be set as	
\begin{equation}
D_0^{\mu \nu} (q)= - \dfrac{\imath T^{\mu \nu} (q)}{q^{2} - M_V^{2}} + \dfrac{\imath L^{\mu \nu} (q)}{M_V^{2}},
\label{propwcero}
\end{equation}
where $T^{\mu \nu} (q) \equiv g^{\mu \nu} - q^{\mu} q^{\nu} /q^{2}$ and $L^{\mu \nu} (q) \equiv q^{\mu} q^{\nu} /q^{2}$, are the  transversal and longitudinal projectors, respectively. The vertex for the process $V(q_1) \rightarrow V(q_2) \gamma(k)$ is defined from the electromagnetic current
\begin{equation}
<V(q_2)|J^\mu(0)| V(q_1)>= A_\nu^\dagger(q_2)A_\lambda(q_1) \Gamma^{\mu \nu \lambda}.
\label{current}
\end{equation}
 At tree level, the vertex can be set as that given by the standard model for the $W$ boson:
\begin{equation}
\Gamma_0^{\mu \nu \lambda} = g^{\mu \nu} (q_1 + q_2)^{\lambda} - g^{\mu \lambda} (q_1 + k)^{\nu} - g^{\nu \lambda} (q_2 - k)^{\mu}  ;
\label{verticewwfcero}
\end{equation}
These expressions satisfy the Ward identity
\begin{equation}
k_\mu \Gamma_0^{\mu \nu \lambda} = \left[\imath D_0^{\nu \lambda} (q_1) \right] ^{-1} - \left[ \imath D_0^{\nu \lambda} (q_2) \right] ^{-1} ;
\label{wardide}
\end{equation}

Upon the inclusion of the finite width of the boson, by considering the loop contributions,  the propagator is modified in a generic form as:
\begin{equation}
D^{\mu \nu} (q) = - \dfrac{\imath T^{\mu \nu} (q)}{q^{2} - M_V^{2} + \imath Im \Pi^{T} (q^{2})} + \dfrac{\imath L^{\mu \nu} (q)}{M_V^{2} -\imath Im \Pi^{L} (q^{2})},
\label{punlazo}
\end{equation} 
where $Im \Pi^{T} (q^{2})$ and $Im \Pi^{L} (q^{2})$, are the transverse and longitudinal part of the absortive contribution of the self-energy induced by the particles in the loop.
Similarly, the vertex becomes
\begin{equation}
\imath e \Gamma^{\mu \nu \lambda} = \imath e (\Gamma_0^{\mu \nu \lambda} + \Gamma_1^{\mu \nu \lambda}), 
\label{elverticecor}
\end{equation}
where $\Gamma_1^{\mu \nu \lambda}$ contains the loop corrections. The Ward identity relates the loop contributions by requiring to satisfy
\begin{equation}
k_\mu \Gamma_1^{\mu \nu \lambda} = \imath Im \Pi^{\nu\lambda} (q_1) - \imath Im \Pi^{\nu\lambda} (q_2).
\label{niw}
\end{equation}

For a boson like the $W$, the scheme consider that  such loops are produced by fermions, while for vector mesons, like the $\rho$, bosons are the natural particles in the loop.

In general, the CP conserving electromagnetic vertex can be decomposed into the following Lorentz structure 
\begin{equation}
\Gamma^{\mu\nu \lambda} = \alpha(k^{2}) g^{\nu \lambda}(q_1 + q_2)^{\mu} + \beta(k^{2}) ( g^{\mu \nu} k^{\lambda} -  g^{\mu \lambda} k^{\nu}) - \gamma(k^{2}) (q_1 + q_2)^{\mu} k^{\nu} k^{\lambda} ,
\label{vercompletomul}
\end{equation}

where the electromagnetic form factors  can be identified as: $|\mathcal{Q}| \equiv \alpha(k^{2})$ is the electric charge form factor ( in $e$ units), $|\vec{\mu}|=\beta(k^{2})\equiv1+\kappa+\lambda$ is the magnetic dipole moment form factor (in $e/2 M_V$ units)  and the electric quadrupole form factor is $|X_E| = \kappa-\gamma(k^2) M_V^2 \equiv \kappa-\lambda$ (in $e/M_V^{2}$ units). The parameters $\kappa $ and $\lambda$ are of common use in the literature to refer to the electromagnetic multipoles  \cite{hagiwara,nieves}. 
The static electromagnetic properties of a particle are defined for the case when the particle is on-shell and in the limit of $\vec{k} \rightarrow 0$.
At tree level, for example, the standard model predicts for the $W$ to have $\alpha(0) = 1$, $\beta(0) = 2$ and  $\gamma(0) =0 $ ($\kappa=1$ and $\lambda= 0$), corresponding to   $|\mathcal{Q}| =1$, $|\vec{\mu}|=2$ and $|X_E| = 1$. Deviations from these values are generically called anomalous and are produced by the inclusion of higher order contributions \cite{Napsuciale:2007ry}. In the present case such contributions are exclusively those required to maintain the electromagnetic gauge invariance, upon the introduction of the finite decay width.

\subsection{$W$ boson form factors}
Let us identify the modification to the $W$ boson form factors introduced by the  correction to the electromagnetic vertex. For that purpouse we consider the explicit expression of the vertex obtained in ref.  \cite{beuthe}, where the mass  of the emitting particles in the loop ($m$) and its weak partner ($m'$) have been considered. The Lorentz structure, transversality and on-shell condition of the boson along with the proper limit for $k \rightarrow 0$ leads to the following expressions for the form factors, defined in equation (\ref{vercompletomul}):

\begin{itemize}
\item Electric charge
\end{itemize} 
\begin{equation}
\alpha(0) = 1 + \imath \sum_i Q_i \frac{\Gamma_i}{M_W}
\left(1
-3\frac{     M_W^2 \Sigma^2(3\Sigma^2- 2M_W^2 )+ \Delta^4(M_W^2-2 \Sigma^2)}{2M_W^6 \lambda^2} 
\right)
\label{fce}
\end{equation}

where $\Sigma^{2}\equiv m^{2} + m^{'2}$, $\Delta^{2} \equiv m^{2} - m^{'2}$ and $Q_i$  is the electric charge of the radiating particle in the loop, 
$\Gamma_i /M_W\equiv g_i^2  \lambda^{3/2}/48\pi$ is the partial decay width for the modes corresponding to the particles in the loop, $g_i$ denotes the coupling and  $\lambda \equiv (M_W^4+\Delta^4-2\Sigma^2 M_W^2)/M_W^4 $ . A sum over all the allowed flavors and color degeneracies is explicitly included. Since the schemes consider the particles in the loop to be on-shell, the flavors include all the leptons and the $u$, $d$, $s$ and $c$ quarks. 

\begin{itemize}
\item Magnetic dipole moment
\end{itemize}
 
  \begin{eqnarray}
\beta(0) = 2 + i \sum_i Q_i \frac{\Gamma_i}{M_W}
\left(
2-
3\frac{\Sigma^2(6\Sigma^2-4M_W^2 )
+\Delta^2(\Sigma^2+3\Delta^2)
+\frac{\Delta^4}{M_W^4}( 2\Delta^4-M_W^2(\Delta^2+7\Sigma^2)}
 {2M_W^4 \lambda^2}
\right)
\label{fdm}
\end{eqnarray}

\begin{itemize}
\item Electric quadrupole moment
\end{itemize}

\begin{equation}
\gamma_\rho(0) = -i \sum_i  Q_i\frac{\Gamma_i}{M_W}   
\frac{ M_W^2 \lambda(2 \Delta^4 - M_W^2(\Sigma^2+\Delta^2)) +    ( \Sigma^2-\Delta ^2)(\Delta^4- M_W^2\Sigma^2)}
{ M_W^6   \lambda^2}
\label{fcue}
\end{equation}

\begin{itemize}
\item \textbf{Isospin limit}
Let us show, just for illustration, the above expressions in the case of $m=m'$. They become
\begin{eqnarray} 
\alpha(0)=1+ \imath \sum_i Q_i \frac{\Gamma_i}{M_W}
\left( 1-   \frac{ 3 \Sigma^2( 3\Sigma^2 -2 M_W^2)}{2 M_W^4 \lambda_I^2} \right),
\end{eqnarray}

\begin{eqnarray} 
\beta(0)=2+ 2\imath \sum_i Q_i \frac{\Gamma _i}{M_W} 
\left( 1-  \frac{ 3 \Sigma^2( 3\Sigma^2 -2 M_W^2)}{2M_W^4 \lambda_I^2 } \right),
\end{eqnarray}

\begin{eqnarray} 
\gamma(0)=\imath \sum_i Q_i \frac{\Gamma _i}{M_W} 
\frac{  \Sigma^2(M_W^2 \lambda_I+\Sigma^2) }
{ M_W^4\lambda_I^2},
\end{eqnarray}
	
\end{itemize}
where $\lambda_I=(M_W^4-2\Sigma^2 M_W^2)/M_W^4 $.

\subsection{$\rho$ and $K^*$ Mesons}
Proceeding along the same lines of the $W$ gauge boson, we obtain the modifications to the $\rho$ meson form factors, taking the complete expressions for the modified vertex computed in \cite{bls}:

\begin{itemize}
\item Electric charge
\end{itemize}
\begin{eqnarray}
\alpha_\rho(0) &=& 1 + i\frac{\Gamma _\rho}{M_\rho}  \left(1- \frac{  3(\Delta^4-\Sigma^2 M_\rho^2)}
{M_\rho^4 \lambda }
\right),
\label{cer}
\end{eqnarray}

where,  $\Gamma _\rho/M_\rho \equiv g^{2}  \lambda^{3/2}/48 \pi $,  $g$ is the effective $\rho \pi\pi$ coupling constant , and $\Delta^2 \equiv m_\pi^{2}-m_{\pi'}^{2}$ and  $\Sigma^{2} \equiv m_\pi^{2}+m_{\pi'}^{2}$.

\begin{itemize}
\item Magnetic dipole moment
\end{itemize}

\begin{eqnarray}
\beta_\rho(0) &=& 2 + i \frac{\Gamma _\rho}{M_\rho} \left( 2+  \frac{3   ( \Sigma^2-\Delta ^2) }{M_\rho^2   \lambda }\right)
\label{dmr}
\end{eqnarray}

\begin{itemize}
\item Electric quadrupole moment
\end{itemize}

\begin{equation}
\gamma_\rho(0) = i \frac{\Gamma _\rho}{M_\rho} 
\frac{ M_\rho^2 \lambda(2 \Delta^4 - M_\rho^2(\Sigma^2+\Delta^2)) +    ( \Sigma^2-\Delta ^2)(\Delta^4- M_\rho^2\Sigma^2)}
{ M_\rho^6   \lambda^2}
\label{cuer}
\end{equation}

\begin{itemize}
\item \textbf{Isospin limit}
The neutral and charged pions are almost degenerated. Taking the isospin limit the form factors become
\begin{eqnarray}
\alpha_\rho(0) = 1 + i \frac{\Gamma_\rho} {M_\rho}  \left(1+\frac{3\Sigma^2}{M_\rho^2 \lambda_I  }\right)
\end{eqnarray}

\begin{eqnarray}
\beta_\rho(0) = 2 + i \frac{\Gamma_\rho} {M_\rho}  \left(2+\frac{3\Sigma^2}{M_\rho^2 \lambda_I  }\right)
\end{eqnarray}

\begin{eqnarray}
\gamma_\rho(0) = -i\frac{\Gamma_\rho}{M_\rho} \frac{  \Sigma^2(M_\rho^2 \lambda_I+\Sigma^2) }
{ M_\rho^4\lambda_I^2}
\end{eqnarray}

\end{itemize}

where $\lambda_I=(M_\rho^4-2\Sigma^2 M_\rho^2)/M_\rho^4 $.

The corrections for the $K^{*+}$ meson follows from the results for the $\rho$ meson by including the two possible channels for the loop contributions: $K^{*+} \rightarrow K^{+} \pi^{0}$ and $K^{*+} \rightarrow K^{0} \pi^{+}$, with the corresponding masses and partial decay widths. 

\subsection{Chiral limit}
The chiral limit correction to the vertex is known to be proportional to the tree level, in both the Fermion and boson loops corrections \cite{bls}. Therefore, in this limit we can write the modification to the form factors in a generic form as follows:
\begin{eqnarray}
\Gamma^{\mu \nu \lambda} = (1+ \imath \frac{\Gamma}{M_V} ) \Gamma_0^{\mu \nu \lambda}
\label{chiral}
\end{eqnarray}

\section{ Numerical results}
The correction to the vertex seems to induce a modification to the the electric charge. However, since the Ward identity is fulfilled, the modification to the vertex is followed by a modification to the propagator, which produces an exact cancelation of the correction to the electric charge. Let us illustrate this point in more detail, for the sake of clarity we consider the expression in the chiral limit:
The modified propagator can be set as \cite{bls,bardin}:
\begin{eqnarray}
D^{\mu \nu} (q) &=& \frac{i}{(1+i\gamma)} 
\left( 
\frac{ -g^{\mu \nu}+ 
\frac{q^\mu q^\nu}{\bar M_V^2-i\bar M_V\bar \Gamma}  }
{ q^{2} - \bar M_V^{2} + i \bar M_V\bar \Gamma}  
 \right)\\
\label{propagadorxral}
&\equiv&<0|T A^{\prime\mu\dagger}(x) A^{\prime\nu}(y)|0>,\nonumber
\end{eqnarray} 
where $\bar M_V, \bar \Gamma \equiv M_V, \Gamma /\sqrt{1+\gamma^2}$, ($\gamma \equiv\Gamma/M_V$),  i.e., it can be seen as a renormalization of the vector field by the inclusion of the finite width 
\begin{equation} 
A^{\prime\nu}=Z^{1/2}A^{\nu}
\end{equation} 
where $Z^{1/2}=1/(1+i\gamma)^{1/2}$. Then, the electromagnetic current becomes
\begin{equation}
A_\nu^{\prime\dagger}(q_2)A^\prime_\lambda(q_1) \Gamma^{\mu \nu \lambda}=
Z A_\nu^\dagger(q_2)A_\lambda(q_1) ( \alpha^\prime(k^{2}) g^{\nu \lambda}(q_1 + q_2)^{\mu} + ...)
\label{current2}
\end{equation}
Therefore, gauge invariance requires that $Z  \alpha^\prime(k^{2})=1$, which is indeed the case, and the electric charge does not receive any correction. Note that the modification to the vertex given by equation \ref{chiral} implies that none of the multipoles receive corrections in the chiral limit.
 An analysis for unstable spin-1/2 particles has also been performed in ref. \cite{scherer}, pointing out to complex renormalization factors as a requirement for properly defined physical quantities. Further considerations on the renormalizability of the wave function can be seen in ref. \cite{renormalization}.\\
The proper values of the modifications to the form factors are then found by the expressions given in the previous section divided by $\alpha^\prime(0)$. In Table \ref{multi}, we present the corresponding results for the $W$ gauge boson and the $\rho$ and $K^*$ mesons. 
Recall that the particles in the loops are restricted to be on-shell. For the $W$  we have included all the leptons and the $u$, $d$, $s$ and $c$ quarks, and the numerical values are similar to those obtained in the chiral limit (the largest $fermion-to-W$  mass ratio is only $\approx 0.022$).	
As a reference of the magnitude of the modification, we recall several results obtained for contributions of different nature: Ref.  \cite{Arbuzov:2009xx} finds a correction to the MDM induced by the Higgs of $\mu = 2 - 0.0151$, reference  \cite{papa}  finds a correction from quarks, leptons and Higgs loops in the SM of $\mu = 2+0.00258$, i.e. the width induced corrections is at least two orders of magnitude smaller than other standard contributions. Note that, due to the renormalization condition, the argument that the finite width correction to the vertex  is of order $\Gamma/M_V$ ($\approx$0.03, 0.19 and 0.06 for the $W$, $\rho$ and $K^*$ respectively) does not extend to the electromagnetic multipoles.

\begin{figure}
\includegraphics{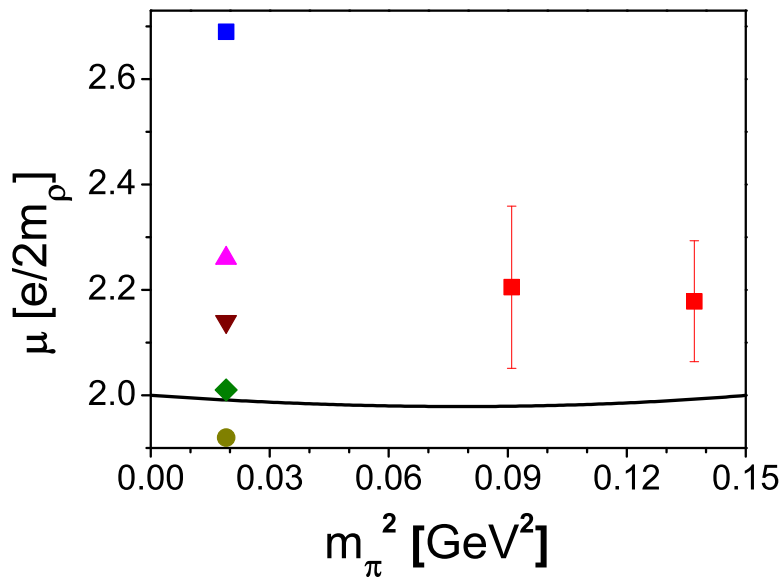}
\caption{\label{dipoloplot} Comparison of the $\rho$ meson MDM as obtained from different approaches. This work (solid line), ref. \cite{hawes} (square), ref. \cite{cardarelli} (triangle up), ref. \cite{melo} (triangle down), ref. \cite{ho} (circle), ref \cite{maris} (diamond) and ref. \cite{hed}   (square with error bars).}
\end{figure}

For the $\rho$ meson, pions are the only on-shell particles allowed in the loops. In this case the $pion-to-\rho$ mass ratio ($\approx 0.18$) is not as small as the corresponding for the $fermion-to-W$, and therefore a significant effect from the mass of the particles in the loop is expected. We can compare our results as shown in Table \ref{multi} with those shown in Table \ref{tabref} as obtained  from several approaches to QCD:  Light-front framework with constituent quarks \cite{melo,cardarelli,ho} and covariant formulations based on the Dyson-Schwinger equations of QCD \cite{hawes,maris}.  In general these last are about the same  or one order of magnitude larger than the finite width induced corrections computed in this work.
Recently, lattice calculations of the form factors have become available  \cite{hed} and, in particular, the dependence on the pion mass they are able to  reproduce has been exhibited. In Figure \ref{dipoloplot}, we compare our result for the MDM with lattice calculations as a function of the pion mass \cite{hed}, we also include predictions from the models at the physical mass of the pion. We observe that the pion mass dependence of our results are mostly flat with a slightly tendency to rise for very large masses. Lattice results  are also flat with a tendency to increase for low masses.

The corrections for the $K^{*+}$ meson is dominated by the $kaon-to-K^*$ mass ratio ($\approx 0.55$) which is very large thus, although the width to mass ratio of the $K^*$ is only $\approx 0.056$, the correction to the multipoles are important.  Compared with the predictions listed in Table \ref{tabref}, they can be  about the same or one order of magnitude smaller.  
 
\begin{table}
\begin{center}
\begin{tabular}{|c|c|c|c|}
\hline
Multipole & $W$ boson & $\rho$ meson & $K^{*}$ meson\\
\hline
 $|\mathcal{Q}|$  [$e$]&1 & 1& 1  \\
\hline
$|\vec{\mu}|$ [$e/2M_V$] & $2.0$ & $2-0.0091$& $2-0.0047$ \\
\hline
 $|X_E|$ [$e/M_V^{2}$] & $1-4.23 \times 10^{-7}$ & $1-0.0387$ & $1-0.097$\\
\hline
\end{tabular}
\end{center}
\caption{$W$ gauge boson, and $\rho$ and $K^*$ mesons multipoles including the correction due to its finite width.}
\label{multi}
\end{table}

\begin{table}
\begin{center}
\begin{tabular}{|c|c|c|}
\hline
Reference & $|\vec{\mu}|$ [$e/2 M_V$] & $|X_E|$ [$e/M_V^2$] \\
\hline
\cite{cardarelli} & $\rho$: 2+0.26 & $\rho$: 1+0.22 \\
\hline
\cite{melo} & $\rho$: 2+0.14 & $\rho$: 1+1.65 \\
\hline
\cite{ho} & $\rho$: 2-0.08 & $\rho$: 1-0.57 \\
\hline
\cite{maris} & $\rho$: 2+0.01 & $\rho$: 1-1.41 \\
\hline
\cite{hawes} & $\rho$: 2+0.69 & $\rho$: 1+1.8 \\
 & $K^{*+}$: 2+0.37 & $K^{*+}$: 1+0.96\\ \cline{2-2}
\hline
\cite{hed} & $\rho$: 2+0.25 & $\rho$: 1-0.75 \\
 & $K^{*+}$: 2+0.14 & $K^{*+}$: 1-0.62\\ \cline{2-2}
\hline
\end{tabular}
\end{center}
\caption{Numerical values for the $\rho$ and $K^*$ mesons  multipoles from several references, computed as stable particles. }
\label{tabref}
\end{table}

As a byproduct, the mean square radius can be computed following \cite{msr}  as, $<R^2>=(\kappa+\lambda)/M_V^2$. For the $\rho$ we obtain a deviation of $-0.0012$ $fm$ respect to the normal value (defined for $\kappa=$1 and $\lambda=0$), which can be compared, for example, with the one computed  in reference \cite{simon}, where they observe a correction of $0.06$ $fm$, due to the inclusion of the pion contribution respect to a pure quark- antiquark state. For $K^*$ the deviation is $-0.0005$ $fm$.
 
\section{Conclusions}	
The  inclusion of the unstable features of spin-1 particles, without breaking the electromagnetic gauge invariance, induces a non trivial modification to the electromagnetic vertex of the particle. In this work we have  extracted the corresponding modifications to the multipole structure of the $W$ and $vector$ mesons. 
Our numerical results for the $W$ gauge boson multipoles shows no substantial deviations from the stable case. For the $\rho$ and $K^*$ mesons, the mass of the particles in the loop makes a significant effect, pointing out that the unstable nature of the vector mesons can be as relevant as other dynamical effects and should be considered in refinements when accounting for their properties.
The modifications in both the propagator and electromagnetic vertex in combination with the Gauge invariance show that the properly defined form factors can be seen as accompanied by a complex renormalization of the vector fields. 

 The general grounds of the loop schemes for spin-1 particles, to account for the finite decay width in a gauge invariant way, have been invoked to study spin-3/2 particles  \cite{LopezCastro:2000cv}. Since in this case the mass ratio between the unstable particle and the ones in the loop can be very large, further studies are desirable to understand at which extend the finite decay width contributes to the multipoles.

\begin{acknowledgments}
We are  grateful to G. L\'opez Castro, J. Piekarewicz and S. Capstick for very useful observations. We also acknowledge the support of CONACyT,  Mexico.
\end{acknowledgments}

\end{document}